\documentclass[%
 reprint,
 amsmath,amssymb,
 aps,
floatfix,
]{revtex4-2}

\usepackage[dvipsnames]{xcolor}
\usepackage{soul}
\usepackage{float}
\usepackage{graphicx}
\usepackage{dcolumn}
\usepackage{bm}

\usepackage{cancel}

\begin{document} 

\preprint{APS/123-QED}

\title{Automatic virtual gate extraction of a 2$\times$2 array of quantum dots\\ with machine learning}


\author{G. A. Oakes}
 \email{gao23@cam.ac.uk}
\affiliation{Cavendish Laboratory, University of Cambridge, J.J. Thomson Avenue, Cambridge CB3 0HE, UK}
\affiliation{Hitachi Cambridge Laboratory, J. J. Thomson Avenue, Cambridge, CB3 0HE, UK}
\author{J. Duan}
\affiliation{London Centre for Nanotechnology, University College London, London WC1H 0AH, UK}
\affiliation{Quantum Motion Technologies, 9 Sterling Way, London, N7 9HJ, UK}
\author{J.J.L. Morton}
\affiliation{London Centre for Nanotechnology, University College London, London WC1H 0AH, UK}
\affiliation{Quantum Motion Technologies, 9 Sterling Way, London, N7 9HJ, UK}
\author{A. Lee}
\affiliation{Cavendish Laboratory, University of Cambridge, J.J. Thomson Avenue, Cambridge CB3 0HE, UK}
\author{C. G. Smith}
\affiliation{Cavendish Laboratory, University of Cambridge, J.J. Thomson Avenue, Cambridge CB3 0HE, UK}
\author{M. F. Gonzalez Zalba}
\email{mg507@cam.ac.uk}
\altaffiliation[\\Present address: ]{Quantum Motion Technologies, Windsor House, Cornwall Road, Harrogate, HG1 2PW, UK}
\affiliation{Hitachi Cambridge Laboratory, J. J. Thomson Avenue, Cambridge, CB3 0HE, UK}

\date{\today}

\begin{abstract}
Spin qubits in quantum dots are a compelling platform for fault-tolerant quantum computing due to the potential to fabricate dense two-dimensional arrays with nearest neighbour couplings, a requirement to implement the surface code.
However, due to the proximity of the surface gate electrodes, cross-coupling capacitances can be substantial, making it difficult to control each quantum dot independently. Increasing the number of quantum dots increases the complexity of the calibration process, which becomes impractical to do heuristically. 
Inspired by recent demonstrations of industrial-grade silicon quantum dot bilinear arrays, we develop a theoretical framework to mitigate the effect of cross-capacitances in 2$\times$2 arrays of quantum dots and extend it to 2$\times$N and N$\times$N arrays. The method is based on extracting the gradients in gate-voltage space of different charge transitions in multiple two-dimensional charge stability diagrams to determine the system's virtual gates.
To automate the process, we train an ensemble of regression models to extract the gradients from a Hough transformation of charge stability diagrams and validate the algorithm on simulated and experimental data of a 2$\times$2 quantum dot array. Our method provides a completely automated tool to mitigate cross-capacitance effects in arrays of QDs which could be utilised to study variability in device electrostatics across large arrays.
\end{abstract}

\maketitle


\section{\label{sec:Intro}Introduction}


Spins in semiconductor quantum dots (QDs) such as those in silicon~\cite{maune2012coherent, pla2012single}, and recently germanium~\cite{hendrickx2020four}, are potential platforms for large-scale quantum computers due to their high qubit density and ease of integration with current semiconductor fabrication lines~\cite{maurand2016cmos}.
Universal quantum computers will be subject to physical errors, requiring quantum error correction protocols to be adopted. The surface code~\cite{fowler2012surface} is the most promising implementation of quantum error correction by distributing logical information over a two-dimensional array of physical qubits with nearest neighbour interactions as long as the individual operations reach a fidelity threshold. The surface code achieves topological protection by increasing the number of physical qubits, making scaling up an even greater and more pressing challenge.
An issue with scaling up is the ability to independently control each QD. Mutual and cross-capacitances make it non-trivial especially given the variability between devices~\cite{yang2020quantum}. A way around this is to apply a combination of voltages corresponding to the eigenvectors perpendicular to electronic transitions that set the virtual gate space. This method has already been implemented to heuristically tune a linear array of eight GaAs QDs~\cite{volk2019loading} and nine QDs in silicon~\cite{mills2019shuttling}. However, as the size of the arrays increases, automatic methods to compensate for capacitive cross-coupling will be necessary. \\

Machine learning has gained popularity in automating various aspects of the tuning protocols in double quantum dots (DQDs). Typically, the first step is to find a set of voltages for which a QD forms under each plunger gate. Black box approaches coarse-tune plunger, barrier and gate voltages to a DQD regime~\cite{botzem2018tuning, teske2019machine, kalantre2019machine, darulova2020autonomous}, allowing tuning devices using direct current measurements faster than a human expert~\cite{moon2020machine} across different semiconducting systems~\cite{Severin2021}.
Commonly, the next step is to narrow down the gate-voltage space in which the system is in a DQD regime. 
Convolutional neural networks (CNNs) have been trained to determine the charge state (no QD, single QD or DQD) of the device at a particular set of gate voltages~\cite{kalantre2019machine, zwolak2020autotuning, darulova2020evaluation}.
Another proposed approach is to take random traces in voltage space to classify the charge state of the system~\cite{zwolak2021ray, zwolak2020ray, krause2021estimation}, which has the advantage of vastly reducing the data and acquisition time required. Thirdly, virtual gates are determined from the slopes of the QD charge transition lines in gate voltage space. This step becomes increasingly important as the size of the QD arrays increases. To extract these gradients, Hough transforms have been implemented near an interdot charge transition (ICT) of a DQD~\cite{mills2019computer}, allowing measuring the device in virtual voltage space. 

Finally, before operating the QD system as a qubit, the device is initialised into a known charge configuration, such as a single electron occupancy in each QD. The proposed methods consist of emptying and then re-loading the desired number of electrons, which operate by using ML techniques to detect transition lines within small voltage patches~\cite{czischek2021miniaturizing, durrer2020automated}. The protocols developed thus far have proved that automated tuning of such complex systems via machine learning is possible. However, as the size of the QD arrays increases, redefined methods that can cope with the increasing parameter complexity need to be developed. \\

Looking at scalable near-term devices, one-dimensional QD  arrays are easier to fabricate~\cite{mills2019shuttling, Philips2022} and should enable the implementation of a logical qubit in a chain of 14 QDs, each hosting a spin qubit~\cite{jones2018logical}. 
The performance of such a logical qubit can be improved by using spins in silicon due to their long-coherence time~\cite{muhonen2014storing, veldhorst2014addressable} and high-fidelity gate set~\cite{yoneda2018quantum, yang2019silicon, Xue2022, Noiri2022}. Furthermore, silicon qubits can be manufactured in industry environments allowing for minimised variability~\cite{maurand2016cmos, Zwerver2022, Camenzind2022}, integrated with local electronics~\cite{morton2011embracing, schaal2017conditional, schaal2019cmos, pauka2019cryogenic, pauka2020characterizing, Ruffino2022} and operated above 1K~\cite{yang2020operation, petit2020universal}, allowing for greater cooling power. More recently, bilinear arrays of silicon quantum dots based on industry-fabricated split-gate transistors \cite{lundberg2020spin, Betz2016, ansaloni2020single, chanrion2020charge, Gilbert2020} have been produced, which should provide additional resources for computing due to their increased connectivity~\cite{hutin2019gate}. Specifically, by using one chain of the array for local radiofrequency (rf) readout via spin-projection of the mirror chain~\cite{gonzalez2015probing,pakkiam2018single, crippa2019gate, west2019gate, zheng2019rapid}, the required depth of the algorithms in Ref.~\cite{jones2018logical} should be reduced. For this reason, the problem of accurately controlling and managing cross-couplings in bilinear QD arrays becomes technologically relevant. \\

In this work, we present an automated method to extract virtual gates to mitigate gate cross-coupling in 2$\times$2 QD arrays. We further show how the method can be extended to 2$\times$N and N$\times$N arrays. We derive an analytical solution of the transformation matrix between real and virtual voltages for the increased connectivity space of a 2$\times$N and N$\times$N arrays of QDs. The equations depend on the gradients in gate-voltage space of regions of charge bistability that can be effectively measured using radio-frequency reflectometry~\cite{Vigneau2022}. We outline a series of measurements to obtain all the required gradients using $in-situ$ dispersive readout (assuming tunnel rates are faster or comparable to the probing radio frequency) and dispersive charge sensing. To automate the process, we have developed an algorithm based on Hough transforms that extracts the required gradients from a stability diagram when sweeping two gate voltages at a time. To manage the effect of non-idealities in real data sets, we train an ensemble of regression models to correctly extract the gradients from the $\theta$ histogram (where $\theta$ is the set of angles between the gate voltage x-axis and the normal to the charge transitions) acquired from the Hough transform of the data set. Finally, we test the trained model on experimental data of a 2$\times$2 QD array and extract partial virtual gates.

\section{\label{sec: Device} Device architecture} 

\begin{figure}[!ht]
    \centering
    \includegraphics[width = \linewidth]{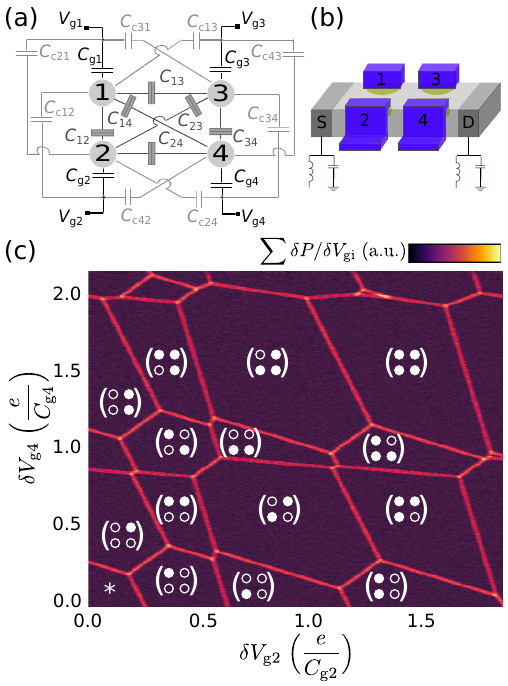}
    \caption{2x2 QD arrays. a) Network of capacitors of a 2$\times$2 QD device consisting of gate $C_{\text{g}i}$, cross $C_{\text{c}ij}$ (\includegraphics{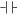}) and mutual capacitors $C_{ij}$. The latter are part of the tunnel barriers (\includegraphics{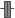}). To simplify the diagram, the cross-capacitors between dots that are diagonally across, as well as the tunnel barriers to reservoirs, have been omitted, but are considered in the model. b) Schematic of a 2$\times$2 QD device consisting of a nanowire with two pairs of split gates. The QDs form at the topmost corners of the nanowire and are defined by the field effect produced by the top gates (depicted in blue). As a result, the device's geometry sets the gate, cross and mutual capacitances. The source (S) and drain (D) of the nanowire act as charge reservoirs to which $LC$ tank circuits are placed for rf reflectometry. c) Theoretical stability diagram whilst as a function of voltages on QDs 2 and 4 ($V_{\text{g1}}$ and $V_{\text{g3}}$ are set to zero). The z-axis represents the change in charge probability ($P$) due to a change in gate voltage. The capacitances used to model the stability diagram can be found in Appendix~\ref{app: capacitance}. The reference charge * on the bottom left of the stability diagram corresponds to $\scriptsize{\begin{pmatrix}\circ & \circ\\ \circ & \circ\end{pmatrix}\equiv\begin{pmatrix}0 & 0\\ 8 & 5\end{pmatrix}}$, to which the addition of a single electron to the corresponding QD and is denoted as $\bullet$.}
    \label{fig:device}
\end{figure}

 In this work, we focus on a 2$\times$2 QD array (see Fig.~\ref{fig:device}a) since it acts as the basic unit for our virtual voltage extraction in 2$\times$N and later in N$\times$N arrays. The electrostatic interaction between QDs can be modelled in the constant interaction (CI) approximation as a network of capacitors: consisting of gate $C_{\text{g}i}$, cross $C_{\text{c}ij}$ and mutual $C_{ij}$ capacitors that are part of the tunnel barriers. An example of a device with such characteristics can be found in a fully-depleted silicon-on-insulator nanowire wrapped by split gates~\cite{ansaloni2020single}, as shown in Fig.~\ref{fig:device}b. QDs form on the topmost corners of the nanowire by applying a voltage on the split gates (depicted in blue). In this architecture, there are no gates to control primarily the tunnel barriers. Instead, the cross and mutual capacitances are predetermined by the physical dimension of the gates and the width of the nanowire. If barrier gates were physically present, they would need to be tuned beforehand, for which  automatic tuning protocols already exist~\cite{moon2020machine, hsiao2020efficient}.
Such device architecture results in higher connectivity than a 1D array, giving rise to a complex stability diagram, as highlighted for a 2$\times$2 array in Fig.~\ref{fig:device}c.

The virtual voltage extraction procedure that we propose relies on being able to detect dot-to-reservoir charge transitions for different two-dimensional stability diagrams. To read every charge transition efficiently, we envisage that $LC$ resonators are connected to the ohmics for dispersive readout~\cite{pakkiam2018single, west2019gate, zheng2019rapid, ibberson2020large} (see Fig.~\ref{fig:device}). Then the resonators can be measured simultaneously using frequency multiplexing.~\cite{hornibrook2014frequency}.
To further increase the signal-to-noise ratio by adding the signals together. In the case where the charge tunnel rates are comparable to or faster than the resonator frequencies, all electronic transitions will be measurable. 
Due to the high sensitivity and bandwidth provided by rf reflectometry techniques, it is possible to measure multiple 2D projections of such a hyper-dimensional voltage space of 300$\times$300 data points in approximately 30 ms~\cite{stehlik2015fast, ibberson2020large}. Thus, by probing two QDs $i$ and $j$ at a time (which we indicate as a superscript) and correctly extracting the gradients $r^{ij}_{i}$ and $r^{ij}_{j}$ of dot to reservoir transitions of QDs $i$ and $j$ (subscript), it is possible to construct the transformation matrix $\bm{G}$ to move from gate-voltage space $\vec{V}_g$ to virtual voltage space $\vec{U}$. In Section~\ref{sec: Background}, we derive the analytical solution for $\bm{G}$ in terms of the different gradients. To automate the process, we develop a protocol to extract the gradients from a stability diagram when sweeping two gate voltages at a time, as outlined in Section~\ref{sec:exp}.

\section{\label{sec: Background} Virtual voltage space}

To model the QD array and gain deeper insight into the effect of cross-capacitive coupling in bilinear and 2D arrays, we generalise the Coulomb blockade model outlined by \textit{Van der Wiel et all.}~\cite{van2002electron}. By treating the array as a network of capacitors, the energy $E$ of a particular charge state is given by:
\begin{equation}
    E = \frac{1}{2} \vec{Q} \cdot \bm{C}^{-1} \vec{Q},
\end{equation}
where $\bm{C}$ is the capacitance matrix, which contains the total capacitance of QD $i$ ($C_{i}$)  as the main diagonal elements, and the mutual capacitances ($-C_{ij}$) in the off-diagonal. The latter corresponds to the interdot capacitances, see Fig.~\ref{fig:device}(a). The charge on each QD, $\vec{Q}$, is then given by:
\begin{equation}
    \vec{Q} = \bm{C_\text{c}} \cdot \vec{V} - |e|\vec{N}, 
\end{equation}
where $\bm{C_\text{c}}$ is the cross-capacitance matrix, with the gate capacitances ($C_{\text{g}i}$) along the main diagonal and the cross-capacitances ($C_{\text{c}ij}$) elsewhere. $\vec{V}$ is the voltage applied onto each QD, and $\vec{N}$ is the number of electrons.
As a result, an analytical solution for any transition between two QDs can be found by equating the two energy states of interest.
Possible transitions are from dot to reservoir (DTR), which have a negative slope or inter-dot charge transitions (ICT), which have a positive gradient. 
By only applying voltages onto two of the QDs ($i$ and $j$) at a time, we take a two-dimensional projection of a hyper-dimensional gate-voltage space. For this specific regime, the gradients $r_{kl}^{ij}$ for a particular transition between two QDs ($k$ and $l$) are given using the Einstein summation convention as:
\begin{eqnarray}
\text{DTR gradient:} \quad \quad r_k^{ij}&&=-\frac{a_{km}C_{\text{c}mi}}{a_{km}C_{\text{c}mj}}\\
\text{ICT gradient:} \quad \quad r_{kl}^{ij} &&= \frac{(a_{km} - a_{lm})C_{\text{c}mi}}{(a_{km} - a_{lm})C_{\text{c}mj}} \nonumber,
\end{eqnarray}
where $a_{ij}$ is the $i, j$ element of $\bm{C}^{-1}$.
In order for the DTR transitions to become orthogonal, we define a new set of axes, which consist of the lines perpendicular to the DTR gradients, as defined by the transformation matrix $\bm{G}$:

\begin{equation}
    \bm{G} = 
    \begin{pmatrix}
        1 & -\frac{1}{r_1^{12}} \\
        -r_2^{12} & 1
    \end{pmatrix}
\end{equation}

\begin{figure}
    \centering
    \includegraphics[width = \linewidth]{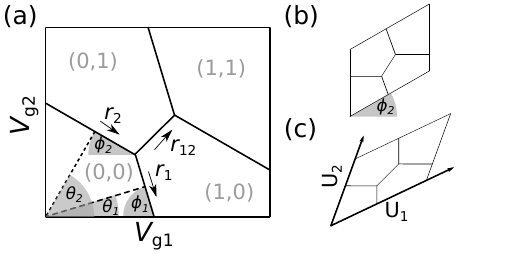}
    \caption{Representation of transformation required to change between gate-voltage space (a) to virtual voltage space (c). This consists of a vertical (b) followed by a horizontal shear (c), such that $r_1$ and $r_2$ are perpendicular to each other.}
    \label{fig:rotation}
\end{figure}

As highlighted in Fig.~\ref{fig:rotation}, this transformation matrix is equivalent to a vertical, followed by a horizontal shear, as well as a re-scaling transformation to ensure that the entries along the main diagonal are unitary, as follows: 
\begin{equation}
    \bm{G} = \begin{pmatrix}
        \frac{r_1-r_2}{r_1} & 0 \\
        0 & 1
    \end{pmatrix}
    \begin{pmatrix}
        1 & -\frac{1}{r_1-r_2} \\
        0 & 1
    \end{pmatrix}
    \begin{pmatrix}
        1 & 0 \\
        -r_2 & 1
    \end{pmatrix}.
\end{equation}

We can extend this procedure to a 2$\times$2 QD array, for which $\bm{G}$ is then given by:

\begin{equation}
\label{eq:transform_2x2}
    \bm{G} = 
    \begin{pmatrix}
        1 & -\frac{1}{r_1^{12}} & -\frac{1}{r_1^{13}} & -\frac{1}{r_1^{14}} \\
        -r_2^{12} & 1 & -\frac{1}{r_2^{23}} & -\frac{1}{r_2^{24}} \\
        -r_3^{13} & -r_3^{23} & 1 & -\frac{1}{r_3^{34}} \\
        -r_4^{14} & -r_4^{24} & -r_4^{34} & 1
    \end{pmatrix}.
\end{equation}


In general, for an N$\times$N array of QDs, the $\bm{G}$ matrix can be constructed as follows:

\begin{equation}
    G_{i,i} = 1 \quad \quad G_{i+n, i} = - \frac{1}{r_i^{i, i+n}} \quad \quad G_{i-n, i} = -r_i^{i-n, i},
\end{equation}

\noindent where $n$ is an integer. For a 2$\times$N array, if only the nearest neighbours are taken into account, $n$ will have values between 1 and 3.

\section{\label{sec:exp} Automatic gradient extraction}

\begin{figure}[!ht]
    \centering
    \includegraphics[width = \linewidth]{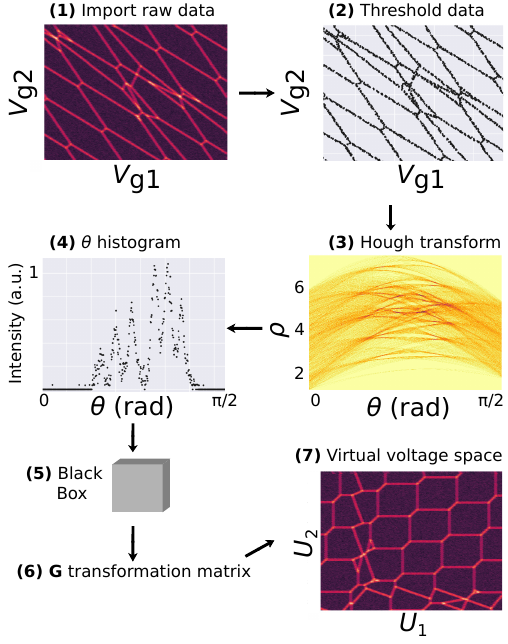}
    \caption{The sequence of steps carried out during automatic gradient extraction protocol. \textbf{1)} Import raw data of stability diagram. \textbf{2)} Threshold the data as described in Appendix~\ref{app: Threshold}. \textbf{3)} Data points are converted into Hough space according to Eq.~\ref{eq:Hough}. \textbf{4)} Histogram of $\theta$ values is extracted and fed to the model in step \textbf{5}. The predicted gradients are then used to extract the partial transformation matrix $\bm{G}$ (step \textbf{6}) that allows partial rotation and shear in virtual voltage space (step \textbf{7}).}
    \label{fig:algorithm}
\end{figure}

We present an algorithm to automatically extract the gradients and thus acquire the virtual voltages from a charge stability diagram when varying two gate voltages.
The protocol is described in Fig.~\ref{fig:algorithm}, and the code can be found here~\cite{GitHub}.
Hough transforms are commonly used in image processing to extract gradients and have already been applied to experimental stability diagrams of DQDs~\cite{mills2019computer, lapointe2019algorithm}. We modify the algorithm to work with data points rather than images, to speed up the analysis.
Only data points representing a signal undergo the Hough transform. Hence the first step is to threshold the data as outlined in Appendix~\ref{app: Threshold}.
The threshold data points ($x,y$) are then converted into Hough space via
\begin{equation}
\label{eq:Hough}
    \rho = x \cdot \text{cos}(\theta) + y \cdot \text{sin}(\theta).
\end{equation}
Since we target the DTR transitions corresponding to the QDs being probed, we only search negative slopes by restricting the $\theta$ values between zero and $\pi/2$. The points in Hough space where multiple sinusoidal lines intersect and hence have higher intensity (as shown in step 3 in Fig.~\ref{fig:algorithm}) are directly related to the gradient $m$ and $y$-intercept, $c$, of the best-fit lines [$m = -1/\text{tan}(\theta)$ and $c=\rho/\text{sin}(\theta)$].

Due to the presence of multiple honeycomb-like structures in the stability diagram of a 2$\times$2 array, the intensity histogram of the $\theta$ values (step 4 in Fig.~\ref{fig:algorithm}) will generally be constituted by more angles than the target 2 per stability map. There can be potentially ten different gradients (four DTR and six ICT transitions) for a 2$\times$2 array, plus additional peaks appear corresponding to lines that combine different transitions, for instance, the lines that pass through the centres of the ICTs with charge configuration $(N,M+1)\leftrightarrow (N+1,M)~\forall~M$. Additional peaks can also appear due to systematic noise in the data, such as aliasing effects arising from the voltage step size used. However, the histograms act as a unique fingerprint to a particular combination of gradients for which we train a regression model to identify the relevant angles in the $\theta$ histogram.

We generated a database of 64,000 theoretical $\theta$ histograms, with the corresponding gradients (calculated according to Eq. 3) from simulated stability diagrams of 2$\times$2 QD arrays. To obtain a broad range of stability diagrams, random capacitance matrices were used, with the physical restriction of having, on average, the gate capacitance $C_{\text{g}i}$ ten times larger than the mutual and cross-capacitances. The data was then convoluted with a Gaussian kernel to replicate the thermal broadening of the transition lines. The data was then split 76.5/13.5/10 as train, validation and test data to train an ensemble of four regression models (neural networks and decision trees) to predict the two main DTR gradients, see Appendix~\ref{app: ML}. We use the average prediction of the different models and their standard deviation to estimate the confidence of the predicted values. From the predicted gradients, we can perform a partial transformation in which the transition lines from the probed QDs become perpendicular to each other, as highlighted in the final step  of Fig.~\ref{fig:algorithm}.

\section{Testing the algorithm on experimental data}

\begin{figure}[!ht]
    \centering
    \includegraphics[width = \linewidth]{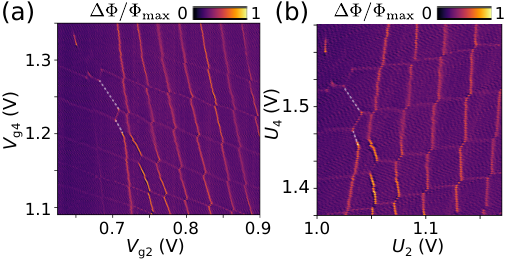}
    \caption{(a) Charge stability map of a 2$\times$2 QD device measured using rf reflectometry. The z-axis represents the normalised phase shift of the reflected signal where high-intensity regions indicate charge bistability (adapted from Ref.~\cite{duan2020remote}). An extra transition line, not originating from QDs 2 and 4, is highlighted by the white dashed line. (b) Data partially transformed into virtual voltage space using $\bm{G}_{i,j\;(i,j \in 2,4)}$ predicted by the model.}
    \label{fig:experiment}
\end{figure}

The regression models presented here have been trained on theoretical data generated using the constant interaction model, which does not consider quantum effects, such as orbital level spacing or voltage and charge occupation-dependent capacitances~\cite{Hanson2007}. To test how it performs on experimental data, we use the charge stability diagram of the left silicon double split-gate nanowire transistors measured in Fig. 1d of Ref.~\cite{duan2020remote}. We present the data in Fig.~\ref{fig:experiment}a, where two gate voltage potentials on the same side of the nanowire ($V_\text{g2}, V_\text{g4}$) are varied while the phase response of the resonator is recorded. We observe quasi-vertical(horizontal) lines of enhanced intensity associated with charge transitions between the QD under gate 2(4) and the reservoirs. 
The DTR transitions from QD 2 are the most intense, as the $LC$ resonator was connected to gate two.
Additionally, we observe a line with a different gradient, highlighted by the dashed white line, that could be originated from charge transitions in the QDs on the opposite side of the nanowire (QD 1 or 3) or from a dopant inside the channel. Either way, the algorithm correctly predicts the main two gradients resulting in the partial rotation matrix:

\begin{equation}
    \bm{G}_{i,j\;(i,j \in 2,4)} = \begin{pmatrix}
    1        & 0.269 \\
    0.479   & 1
    \end{pmatrix}
\end{equation}
As highlighted in Fig.~\ref{fig:experiment}b, the partial rotation predicted by the algorithm results in the DTR transitions of QDs 2 and 4 being perpendicular to each other. To benchmark the accuracy of the transformation, we measure the angles in virtual voltage space and obtain an average angle between DTR transitions of 88.1$\pm$8$^{\circ}$, approaching the desired 90$^{\circ}$.  We note a distribution in the original gradients in Fig.~\ref{fig:experiment}a due to a voltage dependence of the capacitance matrix, highlighting a deviation from the constant interaction model at large voltage amplitudes. We find an average angle between the two DTR transitions in real gate-voltage space of 124$\pm$6$^{\circ}$, whose standard deviation accounts for the majority of the error in the transformation. 
We note that the algorithm can be used to predict virtual voltages locally and to provide average values over wide voltage ranges as described in the next section.

\subsection{\label{sec: Voltage range} Performance at different voltage amplitudes}

\begin{figure}[!ht]
    \centering
    \includegraphics[width = \linewidth]{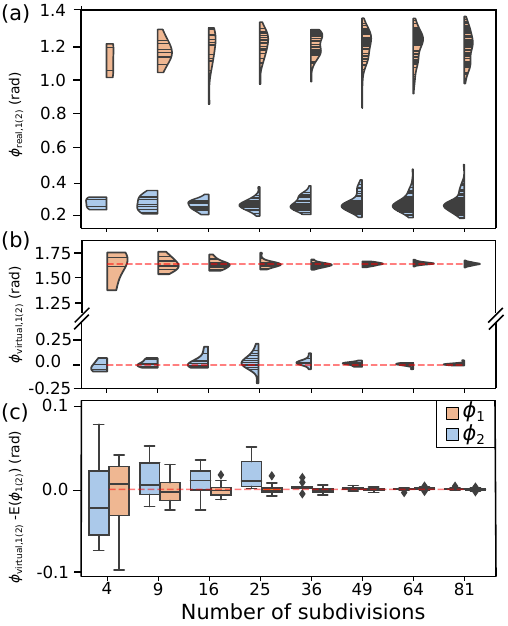}
    \caption{Statistical analysis of the algorithm performance at varying voltage amplitudes by subdividing the experimental data in Fig.~\ref{fig:experiment} into equally sized parts. Violin plots highlighting the distribution of the extracted $\phi_{\text{real},1(2)}$ values in real gate space predicted by the algorithm (a) and the average measured $\phi_{\text{virtual},1(2)}$ gate space after the transformation (b). The red dotted lines target angles 0 and $\pi/2$. (c) Box plot representing the difference between the target ($\frac{\pi}{2}$/0) and measured in virtual voltage space $\phi_{\text{virtual},1(2)}$.}
    \label{fig:subdivide}
\end{figure}

The algorithm is designed to provide the virtual voltages in the voltage range designated by the user. For small amplitudes, the constant interaction model holds, and the algorithm predicts the values across that window. At larger voltage ranges, due to the effect of voltage-varying capacitances in semiconductor QDs, the algorithm provides average values across the desired voltage space, which gives rise to an increased variance in measured angles.   
To assess the accuracy of the algorithm at different voltage scales, we subdivided the data set shown in Fig.~\ref{fig:experiment} into increasingly smaller regions to which we subject the model. We plot the distribution of the predicted gradients in Fig.~\ref{fig:subdivide}a. We then transform the data into the corresponding virtual voltage space according to the predicted gradients. The angles of the DTR lines were manually measured using ImageJ to obtain a statistical analysis of how the model performs (Fig~\ref{fig:subdivide}b).
The process consists of drawing best-fit lines for all the DTR transitions present on each data set to obtain the gradients' average and standard deviation. Since we measure the gradients manually, there may be systematic errors due to inaccuracies in drawing best-fit lines with a mouse. As such, we look more at trends rather than nominal values. We summarise the model's root mean squared error (RMSE) with the increasing number of subdivisions in Table~\ref{tab:acc}. The algorithm accurately predicts the gradients at the different voltage scales studied, showing that it can be used for varying levels of resolution depending on the user's needs. We note the algorithm performs poorly if one of the DTR transitions is not detected. The model then predicts the bias learnt during training (the mean values attained from the training data), namely a $\phi_1 = 1.2\pm0.3$ rad and a $\phi_2 = 0.21\pm0.08$ rad, where $\phi_{1(2)}=\pi/2-\theta_{1(2)}$, as the two angles are complementary to each other (see Fig.~\ref{fig:rotation}). The gradients of the DTRs were omitted from the analysis in such regions, as not enough information is within the data for the algorithm to predict the slopes. We note that the algorithm performs better predicting $\phi_1$ since those transitions have a higher signal-to-noise(SNR) ratio originating from the resonator being connected to gate two. The SNR and data thresholding have an impact on the algorithm's performance.

\begin{table}[ht]
\caption{\label{tab:acc} Root mean squared error of $\phi_{1(2)}$ measured in virtual voltage space as a function of subdivisions.}
    \begin{ruledtabular}
    \begin{tabular}{lcc}
    Subdivisions & RMSE $\phi_1$ (mrad) & RMSE $\phi_2$ (mrad) \\
    \hline
    1    & 88.64 & 107.5 \\
    4    & 1.61 & 116.7 \\
    9    & 7.04 & 93.8 \\
    16   & 1.74 & 95.0 \\
    25   & 0.39 & 80.1 \\
    36   & 0.67 & 19.9 \\
    49   & 0.15 & 7.6 \\
    64   & 0.05 & 3.2 \\
    81   & 0.07 & 1.5  
    \end{tabular}
    \end{ruledtabular}
\end{table}

\section{\label{sec:2x2} Extracting virtual voltages for a 2$\times$2 array of QD\lowercase{s}}

\begin{figure}[!ht]
    \centering
    \includegraphics[width = \linewidth]{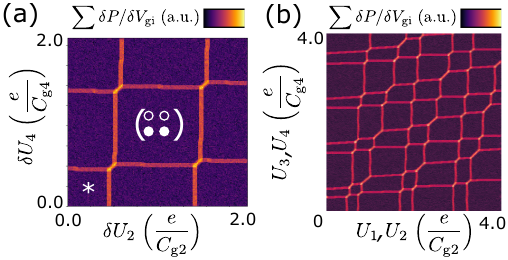}
    \caption{Simulated virtual gate space once the entire $\bm{G}$ matrix is extracted. a) Corresponding virtual gate space of Fig.~\ref{fig:device}c, with a charge state denoted as reference. The reference charge * is identical to that used in Fig.\ref{fig:device}c. b) To show independent control of all QDs, all four virtual voltages are swept, $U_{\text{1}}$ and $U_{\text{2}}$ along the horizontal and $U_{\text{3}}$ and $U_{\text{4}}$ along the vertical axis.}
    \label{fig:virtual}
\end{figure}

Theoretically, ten types of charge transitions are possible for a 2$\times$2 QD array, resulting in ten different gradients that could be experimentally measured when applying voltages to two gates. Several of these are related to each other by:
\begin{equation}
r^{ij}_{kl} = r^{im}_{kl} \cdot r^{mj}_{kl},
\end{equation}
thus, reducing the number of measurements required to define $\bm{G}$ for a 2$\times$N array to 2N-1. However, it is not trivial to correctly label each measured gradient to its corresponding transition. Also, many of these transitions will only appear at particular gate voltage combinations, requiring large gate voltage maps. A more robust solution is to extract only the two gradients of the QDs whose primary gates are being swept and thus take $5\text{N}-4$ stability diagrams whilst covering relatively smaller gate voltage ranges. Although more measurements are taken, they can be smaller range voltage maps, minimising the impact on total measurement time whilst guaranteeing that all the transitions are measured. Another advantage is that we can implement the automatic gradient extraction protocol introduced in Section~\ref{sec:exp} for each two-dimensional projection. 
We performed the virtual voltage extraction protocol on a simulated 2$\times$2 QD array by first simulating the six different 2D gate voltage projections to extract all the relevant gradients.
By constructing the $\bm{G}$ matrix, as in Eq.~\ref{eq:transform_2x2}, we determine the corresponding virtual voltages $\vec{U}$ in terms of gate voltages $\vec{V}_g$. Figure~\ref{fig:virtual}a shows the complete virtual voltage space transformation corresponding to Fig.~\ref{fig:device}c, highlighting how it simplifies the analysis as there are no transitions from QDs 1 and 3. To prove this method enables controlling each QD independently of its neighbours, in Fig.~\ref{fig:virtual}b, we plot the charge stability diagram when simultaneously sweeping the virtual voltages $U_{\text{1,2}}$ and $U_{\text{3,4}}$ in the $x$ and $y$-axis, respectively, after applying the virtual voltage extraction protocol.  

\section{Extending to Larger QD arrays}

So far, our algorithm considers that each QD in the array can exchange charges with reservoirs at a rate similar to or faster than the rf probing frequency, meaning that each DTR can be measured directly via radio-frequency reflectometry~\cite{Vigneau2022}. In Appendix~\ref{app: Direct_architectures}, we present scalable QD arrays (bilinear and sparse two-dimensional) that retain this property, and hence the algorithm can be used as is to construct the $\bm{G}$ transformation matrix. To go beyond this restriction, in this Section, we generalise our procedure to automatically extract the virtual gates in 2$\times$N and N$\times$N arrays when QDs can only exchange charges with reservoirs at rates much slower than the probing frequency. For this purpose, we utilise the concept of dispersive charge sensing via a single-electron box (SEB)~\cite{House2016,oakes2022fast}, where the admittance of a QD tunnel coupled to a reservoir is used for charge sensing purposes. We illustrate the enhanced protocol for a 2$\times$6 array in Fig.~\ref{fig:array}. To probe the admittance of the SEBs, we connect rf resonators on the source and drain ohmic contacts since it minimises the number of $LC$ circuits needed for the protocol. 
We divide the protocol into three steps and focus on the source side to simplify the explanation:
\begin{figure}[!ht]
    \centering
    \includegraphics[width = \linewidth]{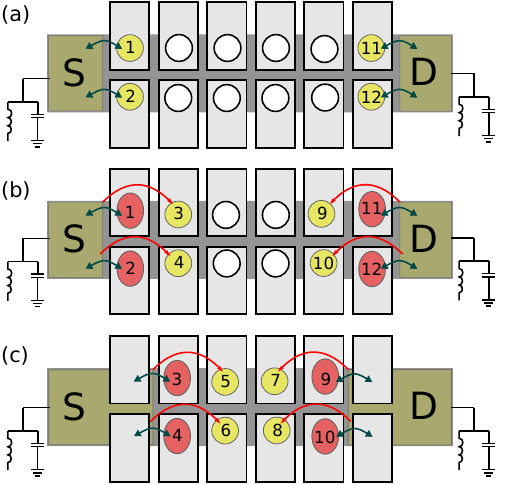}
    \caption{Extending the 2$\times$2 QD array protocol to a 2$\times$N array, with an exemplary procedure for a 2$\times$6 array. To be able to measure every pair of QDs, we place $LC$ resonators at the source and drain. a) We first characterise the outer pairs of QDs, which are directly connected to a reservoir (yellow circles). b) The characterised QDs can then be used as charge sensors (red ellipses), allowing us to measure QDs 3, 4, 9 and 10. c) By applying even higher voltages on the outer QDs, we can extend the reservoirs and thus operate QDs 3, 4, 9 and 10 as charge sensors, allowing us to characterise the innermost QDs.}
    \label{fig:array}
\end{figure}

(i) The first step is measuring the stability diagrams for the pairs of QDs 1-2. The two QDs have direct access to a reservoir and thus can be measured directly using reflectometry, see Fig.~\ref{fig:array}a. This step allows extracting the coefficients of the $\bm{G}$ matrix $G_{12(21)}$, see Appendix~\ref{app: PID}. (ii) The second step focuses on the QDs not directly connected to reservoirs. i.e. 3-4 (Fig.~\ref{fig:array}b). We operate QDs 1-2 as SEBs for charge sensing DTR of QDs 3-4~\cite{oakes2022fast, Hogg2022, niegemann2022}. Operating the SEB as a charge sensor requires voltage compensation to maintain the SEB at a point of charge instability which can be achieved using a proportional integral differential (PID) feedback control system~\cite{Yang2011}. Sweeping pairs of gate voltages of the 1-4 subset with PID feedback results in measuring modified projections of the voltage space than the commonly measured ones when sweeping just two gate voltages. However, the effect of the PID can be compensated when constructing the $\bm{G}$ transformation matrix, as outlined in Appendix~\ref{app: PID}. (iii) The final step focuses on characterising the innermost QDs 5-6, see Fig.~\ref{fig:array}c. Here, we apply high voltages to gates 1-2 to produce a strong inversion, resulting in the extension of the source charge reservoir towards QDs 3-4 which are now used as SEBs. We assume this will have a negligible effect on QDs 3-4 due to their small cross-capacitance to the reservoir, as they are not nearest neighbours and thus can repeat step ii. Once all the required gradients have been extracted, we construct the final transformation matrix $\bm{G}$ and obtain the virtual gates of the array.
\begin{figure}[ht]
    \centering
    \includegraphics[width=\linewidth]{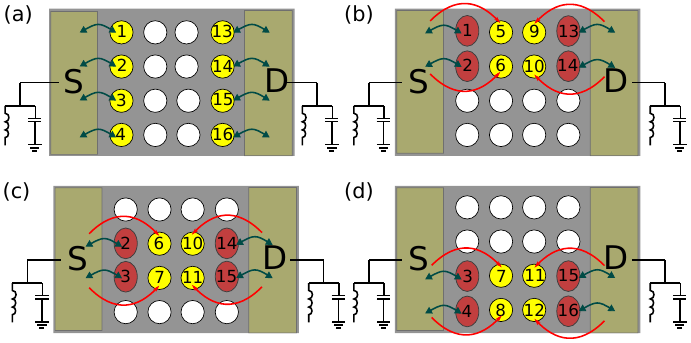}
    \caption{Extending the 2$\times$2 QD array protocol to an N$\times$N array. We assume that $LC$ resonators are connected to the source and drain and a similar protocol to the one used for the 2$\times$N array is used. a) Firstly, pairs of QDs where both have direct access to a reservoir are measured (yellow circles). (b-d) To accurately measure the inner QDs, the outer ones are used as charge sensors (depicted in red).}
    \label{fig:NxN}
\end{figure}

\begin{figure}[ht]
    \centering
    \includegraphics[width=\linewidth]{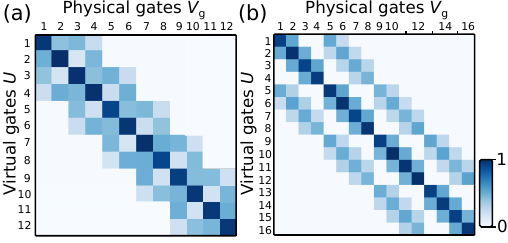}
    \caption{Visualisation of the $\bm{G}$ matrix for a) a 2$\times$6 array, as in Fig.~\ref{fig:array} and b) a 4$\times$4 array, as shown in Fig.~\ref{fig:NxN}. The entries of each row show how the virtual voltage values, and hence the electrochemical potential, of a QD, is influenced by each physical gate. In both cases, we form a sparse matrix hence only stability maps of nearest neighbour QDs need to be measured to construct $\bm{G}$. The difference in the distribution of entries between matrix a) and b) reflects the QD connectivity given by the respective architectures.}
    \label{fig:matrix}
\end{figure}

A similar protocol can be adapted to extract virtual gates in N$\times$N arrays. Here, we consider architectures where the array is sandwiched between a source and drain charge reservoir and the QDs have contact over active gates~\cite{Weinstein2023}. We first characterise every pair of QDs that have direct access to a reservoir (Fig.~\ref{fig:NxN}a) and then form SEB charge sensors to detect charge transitions in QDs in the central part of the structure (Fig.~\ref{fig:NxN}b-d). If needed, the reservoirs can be extended to form SEBs towards more central regions of the structure, similar to step (iii) above. In general, for an N$\times$M array of QDs, one would need to measure $4\text{NM} - 3(\text{N}+\text{M}) + 2$ stability diagrams to reconstruct the whole $\bm{G}$ matrix (where non-nearest neighbours are assumed to have a negligible contribution). The expected $\bm{G}$ matrix entries for a 2$\times$6 and a 4$\times$4 QD array are summarised in Fig.~\ref{fig:matrix}.

The proposed extended protocol alternates between using QDs as charge sensors and data qubits. SEBs can operate as charge sensors for an arbitrary number of charges inside the box, provided that the tunnelling rate to the sensing QD is similar to or faster than the rf probing frequency ~\cite{Vigneau2022}. Therefore, SEBs operated in the few-electron regime will have a similar bias condition to the operation region of the data qubits. Hence, the capacitance matrix will be unaltered between the two regimes. However, if SEBs were operated in the multi-electron regime to enhance the tunnel rates to the reservoirs, the capacitance matrix could alter due to the bias-dependent shape of the QDs. We set bounds to this change using Fig.~\ref{fig:experiment}a. We determine a gradient change of $\sim$2$^{\circ}$ every 0.25~V, from which we estimate a slope difference of $\sim$4 to 8$^{\circ}$ between the multi- and few-electron regimes. We note that these estimated values are technology dependent. From Eq. 3, we can predict which gradients in the low electron regime are steeper or shallower than the ones measured with our protocol. Hence, we can subtract/add the average estimated deviation to the measured gradients to obtain an even better estimate of $\bm{G}$ in the few-electron regime.

\section{Outlook}
In this paper, we have developed the relation between gate voltage and virtual voltage space for a 2$\times$N and an N$\times$N array of QDs in terms of the gradients of the DTR transitions obtained whilst simultaneously probing two QDs. We then propose a protocol of the different measurements required to construct our transformation matrix $\bm{G}$ between the gate and virtual voltage space. To automate the process we developed an algorithm to extract the appropriate gradients. This is done by utilising the Hough transform of a stability diagram when probing two gate voltages at a time. To manage the complexity of Hough's $\theta$ histogram obtained from experimental data, we train an ensemble of regression models to predict the two most prevalent DTR gradients. We then tested our protocol on experimental data of a 2$\times$2 QD array. We outline how the protocol can be extended to a 2$\times$N and N$\times$N arrays of QDs.
Further work is required to incorporate barrier compensation within the protocol, which would generalise our algorithm to more complex geometries. Finally, the virtual gate extraction methodology presented in this Article could be particularly useful in studying the direct and cross-coupling capacitance variability across QDs in large 2$\times$N and N$\times$N arrays.

\begin{acknowledgments}
We thank Benoit Bertrand, Louis Hutin and Maud Vinet for providing the samples and for useful discussions. This research has received funding from the European Union’s Horizon 2020 Research and Innovation Programme under grant agreement No 688539 (http://mos-quito.eu) and  951852, and the Winton Programme for the Physics of Sustainability. G.A.O. acknowledges support from the EPSRC Cambridge NanoDTC, EP/L015978/1. J.D. acknowledges support from the EPSRC CDT in Delivering Quantum technology EP/L015242/1. C.G.S. acknowledges support from the EPSRC grant number EP/S019324/1. M.F.G.Z. acknowledges support from the Royal Society.
\end{acknowledgments}

\section*{Data availability}
The data and code supporting the findings of this study are openly available in the following GitHub repository~\cite{GitHub}.

\appendix
\section{\label{app: Threshold} Threshold experimental data}

\begin{figure}[!ht]f
    \centering
    \includegraphics[width = \linewidth]{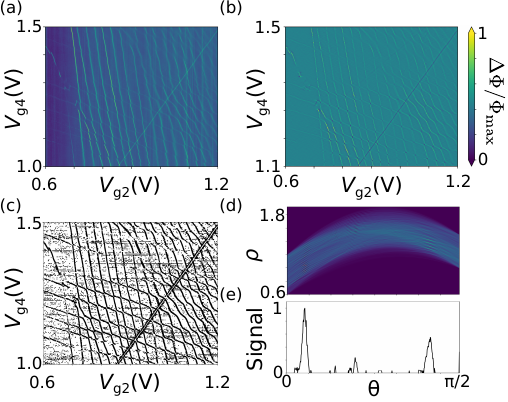}
    \caption{(a) Raw reflectometry data of a 2$\times$2 Si QD device. (b) Data after being flattened with wavelet edge detection. (c) Data above the threshold value determined by the Kullback-Leibler divergence is then Hough transformed in (d), giving rise to the $\theta$ histogram in (e).}
    \label{fig:flatten}
\end{figure}

We have developed an automated thresholding protocol, which does not require any fitting parameters. Rather, it uses information theory to determine the threshold between noise and signal. Experimentally in an rf reflectometry setup, the reflected signal goes through an IQ mixer, allowing the measurement of both the in-phase (I) and quadrature (Q) components. From the measured I and Q, one can calculate the magnitude ($M=\sqrt{I^2+Q^2}$) and phase ($\Phi=\text{tan}^{-1}(Q/I)$), where changes in the phase are proportional to the changes in capacitance of the system ($\Delta C$)~\cite{gonzalez2016gate}. In particular, at regions of charge bistability, an additional capacitance appears, resulting in a phase shift ($\Delta \Phi$) that can be approximated as:
\begin{equation}
    \Delta \Phi \approx \frac{-2Q_\text{L} \Delta C}{C_\text{p}},
\end{equation}
where $Q_\text{L}$ is the loaded quality factor of the resonator and $C_\text{p}$ is the parasitic capacitance of the system. Besides this shift, in gate-defined QD, an additional capacitance associated with the two-dimensional charge inversion layer generated by the gate will also contribute to $\Delta \Phi$~\cite{rossi2017dispersive}. This effect is visible in Fig.~\ref{fig:flatten}a, as the background signal of the charge stability diagram increases as the resonator gate voltage, $V_{\text{g2}}$, is increased. Over large voltage ranges, such a background is typically modelled as a sigmoid function.
The features that we want to detect, however, are characteristic of a faster rate of change in phase shift compared to the sigmoid background. Therefore, inspired by Canny edge detection, we first convoluted the signal with a 1/cosh$^2$ kernel, which is the expected shape of the tunnelling capacitance of a thermally broadened DTR transition~\cite{ahmed2018primary}. This has the effect of smoothing the data from high-frequency noise and promoting any detected peaks. We then use the first derivative of a Gaussian as a mother wavelet for edge detection~\cite{Prance2015}, with ten different scales ranging uniformly on a log scale from five data points to a tenth of the total voltage range. The average spectrum of the wavelet transform gives rise to the flattened and cleaned image in Fig.~\ref{fig:flatten}b.

\begin{figure}[!ht]
    \centering
    \includegraphics[width = \linewidth]{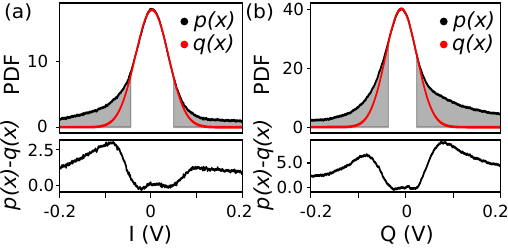}
    \caption{Histograms of the intensities of the  I (a) and Q (b) data sets (black data points). A Gaussian distribution is fitted to the I and Q histograms (red curve). Kullback-Leibler divergence is then used to determine at which point the data set is no longer well approximated by a Gaussian distribution, as highlighted by the residuals in the insert beneath, which represents the threshold point (grey-shaded region). }
    \label{fig:AC_DC}
\end{figure}

The next step is to threshold the data. For most gate voltage combinations the QDs have a particular charge stable configuration, which does not generate a signal. Only in regions of charge bistability, the electrons tunnel cyclically driven by the effect of the rf voltage producing a phase response. Therefore, for most combinations, we do not measure a signal, which only arises at a charge transition. Thus, we expect to have a large peak in the intensity distribution, representing the background, which will follow a Gaussian distribution if the noise is dominated by the HEMT amplifier. To determine what value $x$ to threshold for I and Q, we use the point that minimises the Kullback-Leibler divergence $D_{KL}$ between the square root of the intensity distribution $p(x)$ (black data points in Fig.~\ref{fig:AC_DC}) and that of a fitted Gaussian distribution $q(x)$ (red fitted line in Fig.~\ref{fig:AC_DC}).
By taking the square root of the intensities of the histogram data, it broadens the peak, allowing us to distinguish weak signals in the data~\cite{venderley2020harnessing}. The Kullback-Leibler divergence $D_{KL}$ calculated gives the information loss of approximating the intensity distribution of the data $p(x)$ by a Gaussian $q(x)$, which represents the background:
\begin{equation}
    D_{KL}(p(x)|q(x)) = \sum_{x \in X} p(x) \; ln\left(\frac{p(x)}{q(x)}\right)
\end{equation}
By minimising $D_{KL}$, we have found the data points of $p(x)$ that are best described by $q(x)$ and thus represent our background, while data points that diverge from a Gaussian distribution (shaded in grey in Fig.~\ref{fig:AC_DC}) represent our signal.

The reason for thresholding I and Q separately is that some transitions appear stronger in one domain than in the other. Therefore data points labelled as a signal in either I or Q are taken into account, resulting in Fig.~\ref{fig:flatten}c.  

Even though we observe some random noise in Fig.~\ref{fig:flatten}c, it is not an issue for the Hough transform (in Fig.~\ref{fig:flatten}d), as these data points are unlikely to generate a peak in the $\theta$ histogram due to their random orientation.
From the $\theta$ histogram in Fig.~\ref{fig:flatten}e, we observe two prominent peaks, one close to zero representing the DTR transitions from gate two and the other close to $\pi/2$ due to the DTR transitions from QD four. There is also a peak slightly below $\pi/4$, arising from the single QD region on the top right corner of the stability diagram.

\section{\label{app: ML} Ensemble regression models}

\begin{figure}[ht]
    \centering
    \includegraphics[width = \linewidth]{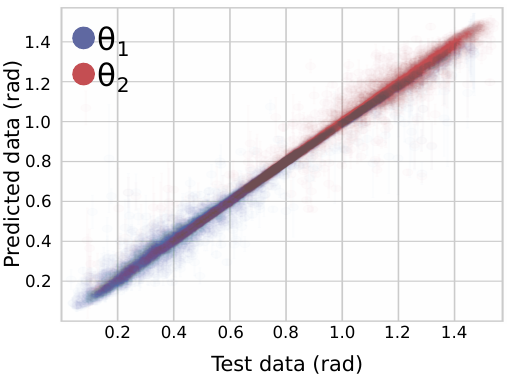}
    \caption{Predicted $\theta$ values by the model versus their true value on the 6,500 test data points. Error bars are also plotted, corresponding to the standard deviation between the predicted values of the four models that make the ensemble. The better the trained model, the closer the data should resemble $y=x$.}
    \label{fig:NN}
\end{figure}

The data set used to train the model was created by generating 64,000 random stability diagrams of a 2$\times$2 QD array. The data set contains examples of single, double and quadruple dots. From the stability diagrams generated, the $\theta$ histograms were extracted. The histograms were standardised by making them 500 data points long, ranging from 0 to $\pi/2$. This ensures that only negative gradient lines are fitted to the data and thus excluding any ICT. The intensities of these histograms were normalised to have a maximum value of one. Since the capacitances used to generate these theoretical stability diagrams are known, from Eq.~3, we can calculate the gradients of the two DTR transitions of interest, which act as the labels for the data set.   
The slopes were converted to their corresponding $\theta$ values and divided by $\pi/2$ to normalise the labels between zero and one.\\
Ensemble regression aims to train different regression models to do the same task. By taking the average predictions: one obtains a more accurate and robust model than any of its constituents. Ensemble models rely on the ``wisdom of the masses," as multiple hypotheses could equally fit the training data. If the trained models are uncorrelated, taking their average reduces the risk of choosing the wrong hypothesis. This is particularly relevant as often, during training, the model can get stuck in a local optimum~\cite{louppe2014understanding}. The standard deviation between the different predictions of the ensemble models is utilised to determine how confident we are of the predicted value.
To ensure that the trained models are uncorrelated, we use four different popular regression models, consisting of a neural network and three decision tree-based algorithms, namely: random forest~\cite{breiman2001random}, bagging regression~\cite{breiman1999pasting} and extra trees regression~\cite{geurts2006extremely}. Out of the four models, the neural network is the one that can model more arbitrary functions but requires more hyperparameters to be tuned during training. For the neural network, as a non-linear transform, a ReLu activation function is used between each layer. He uniform initialisation~\cite{he2015delving} is used to ensure that gradients of the activation function do not vanish to zero. The architecture used consists of 4 hidden layers with 512 nodes each and, between each layer, a 0.05 probability dropout as regularisation to ensure the model does not over-fit. The learning rate is set to $10^{-4}$. The learning curve was used during training to ensure that the model did not over-fit the training data.
The three tree-based models are already ensembles of multiple decision trees. The way they differ is how they randomly select subsets for training and optimise each decision tree, thus giving rise to a different hypothesis.
The trained models were then evaluated on the test data, as shown in Fig.~\ref{fig:NN}, which shows that the model works very well, with an $R^2$ value of 0.978 and an average accuracy for $\theta_1$ of 97.36\% and 99.29\% for $\theta_2$.
For $\bm{G}$, however, we are interested in the gradients $r = -\frac{1}{\text{tan}(\theta)}$, making the conversion of the two trivial.

\section{\label{app: capacitance} Parameters used in simulated data}

We summarise the numerical values used to produce Fig.~\ref{fig:device} and Fig.~\ref{fig:virtual} in Table~\ref{tab:cap},

\begin{table}[!ht]
\caption{\label{tab:cap}Capacitance $C$ and cross-capacitance $C_{\text{c}}$ matrices that generated Fig.~\ref{fig:device}c and Fig.~\ref{fig:virtual} (rounded to four decimal places). The capacitances used have no physical units because we are only interested in their relative ratios, as these will give rise to a particular stability diagram. }
\begin{ruledtabular}
\begin{tabular}{ccccccccc}
 \multicolumn{4}{c}{$C$ matrix} &{\hspace{2em}}& \multicolumn{4}{c}{$C_{\text{c}}$ matrix}\\
 1.6199 & -0.4084 & -0.0662 & -0.0364 &  & 1.0225 & 0.0486 & 0.0272 & 0.0106 \\
 -0.4084 & 1.8513 & -0.0558 & -0.3077 &  & 0.0587 & 0.9519 & 0.0119 & 0.0569 \\
 -0.0662 & -0.0558 & 1.6845 & -0.3806 & & 0.0481 & 0.0322 & 1.0549 & 0.0467 \\
 -0.0364 & -0.3077 & -0.3806 &  1.8772 & & 0.0483 & 0.0287 & 0.0973 & 0.9783 \\
\end{tabular}
\end{ruledtabular}
\end{table}
\noindent

To produce the simulation in virtual gate space in Fig.~\ref{fig:virtual}b , we utilise the following $\bm{G}$ matrix,

\begin{equation}
    \bm{G} = \begin{pmatrix}
            1        & 0.2643 & 0.0991 & 0.0919 \\
            0.3469   & 1 & 0.1308 & 0.2528 \\
            0.1236   & 0.1099 & 1 & 0.2452 \\
            0.1592   & 0.2292 & 0.3498 & 1
    \end{pmatrix}
\end{equation}

\section{\label{app: Direct_architectures}Directly measurable 2$\times$N and N$\times$N array architectures}

\begin{figure}[ht]
    \centering
    \includegraphics[width=\linewidth]{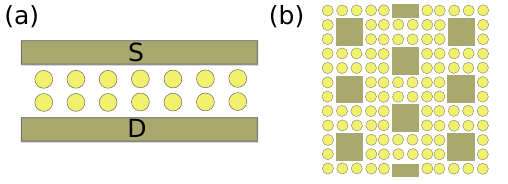}
    \caption{Examples of architectures for which all DTR transitions can be directly measured. a) A 2$\times$N array and b) a sparse two-dimensional array. QDs are represented by yellow circles, while the reservoirs are dark yellow rectangles.}
    \label{fig:tune_direct}
\end{figure}

Here, we present scalable architectures in which every QD has direct access to a reservoir (fast tunnelling condition), thus, every DTR can be measured via direct radio-frequency reflectometry. In Fig.~\ref{fig:tune_direct}a, we highlight how a 2$\times$N array of QDs can be sandwiched between two parallel reservoirs that can act as source and drain. The gates are placed vertically onto the QDs ~\cite{Weinstein2023}.  
The concept can be scaled up to sparse two-dimensional arrays, see Fig.~\ref{fig:tune_direct}b. Here, we consider squared reservoirs surrounded by 12 QDs each, which acts as the motif of a periodic 2D hexagonal lattice. Again, the QDs, as well as the reservoirs are gated vertically. We note that in these types of architectures, multiple 2$\times$2 sub-blocks can be tuned simultaneously using frequency multiplexing techniques to reduce the time to extract all virtual gates.

\section{\label{app: PID}Effect of PID on virtual gate extraction}

In many proposed QD array architectures, not all the QDs have direct access to a reservoir, which means that a DTR transition may not be detectable using direct radio-frequency reflectometry since the fast tunnelling condition may not be met. Thus the protocol presented in Sec.~\ref{sec:exp} needs to be modified. To overcome this challenge, we propose an extension to our algorithm. We use QDs with a direct connection to a reservoir (for which virtual gates can be extracted), as single-electron box (SEB) charge sensors to extract the gradients of the remaining QDs\cite{oakes2022fast,Hogg2022}. We operate the sensor in a Partial Integral Differential (PID) feedback loop~\cite{Yang2011} to maintain the sensitivity of the sensor over wide gate voltage ranges.

The necessity to modify the algorithm comes from the fact that varying the voltage applied on the charge sensor affects the slopes of the stability diagram measured and, thus, the $\bm{G}$ matrix. Here, we outline the relation of the PID measured slopes $r_k^{i,j}\big|_{(l,m)}$, where $l$ and $m$ are QDs operated as charge sensors, to the $\bm{G}$ matrix coefficients. To illustrate the algorithm extension, we concentrate on QDs 1-4 on the left in Fig.~\ref{fig:array}. We assume the cross-capacitive effects from non-nearest-neighbour QDs are negligible. Before operating QDs 1 and 2 as SEBs, we have characterised their cross-capacitance (Fig.~\ref{fig:array}a) resulting in the partial $\bm{G}$ matrix
\begin{equation}
\label{eq:G_direct}
    \bm{G} = 
    \begin{pmatrix}
        1 & G_{12} &  \dots & 0\\
        G_{21} & 1  & & \\
        \vdots &  & \ddots & \\
        0 &  &  & 1 
    \end{pmatrix}.
\end{equation}

We then fill in the remaining entries of Eq.~\ref{eq:G_direct} by moving to the configuration in Fig.~\ref{fig:array}b. We divide the process into two steps: 
(i) We operate the SEB 1(2) and measure the charge stability diagrams varying gates 2-3 and 2-4 (1-3 and 1-4). The measured slopes are
\begin{align}
    \label{eq:r_ij}
    r_i^{ij}\big|_{(l)} =& -\frac{a_{im}C_{\text{c}mi} + G_{li}\left(a_{im}C_{\text{c}ml}\right)}{a_{im}C_{\text{c}mj} + G_{lj}\left(a_{im}C_{\text{c}ml}\right)}\\
    \label{eq:r_ji}
    r_j^{ij}\big|_{(l)} =& -\frac{a_{jm} C_{\text{c}mi} + G_{li}\left(a_{jm}C_{\text{c}ml}\right)}{a_{jm}C_{\text{c}mj} + G_{lj} \left(a_{jm}C_{\text{c}ml}\right)}.
\end{align}

By using Eq.~7 and rearranging Eq.~\ref{eq:r_ij} and~\ref{eq:r_ji}, we get the following relations
\begin{align}
     G_{ij} =& - \frac{1 + r_i^{ij}\big|_{(l)}(G_{il} G_{lj})-  G_{il} G_{li} }{r_i^{ij}\big|_{(l)}} \\
     G_{ji} =& -r_j^{ij}\big|_{(l)}(1-G_{jl}G_{lj}) - G_{jl} G_{li}.
\end{align}

N.B. that $G_{il}=G_{jl}$ and $G_{lj}$ are unknown. However, we end up with eight equations and eight unknowns so that we can extract all the required parameters 
\begin{align}
    G_{ij} =& - \frac{r_l^{lj}\big|_{(i)} - G_{il}r_i^{ij}\big|_{(l)}}{r_i^{ij}\big|_{(l)} r_l^{lj}\big|_{(i)}}\\
    G_{ji} =& - \frac{r_j^{ij}\big|_{(l)} +r_j^{lj}\big|_{(i)}\left(G_{lj}r_j^{ij}\big|_{(l)} -G_{li}  \right)}{1 - \left( G_{lj} r_j^{ij}\big|_{(l)} - G_{li} \right)\left( G_{ij} r_j^{lj}\big|_{(i)} - G_{il} \right)}.
\end{align}

At this point, we have extracted the following entries for our transformation matrix $\bm{G}$
\begin{equation}
\label{eq:G_PID}
    \bm{G} = 
    \begin{pmatrix}
        1 & G_{12} &  G_{13} &  G_{14} & \dots & 0\\
        G_{21} & 1 &  G_{23} &  G_{24} & & \\
        G_{31} & G_{32} & 1 & & &\\
        G_{41} & G_{42} & & 1 & &\\
        \vdots &  & & & \ddots & \\
        0 &  &  & & & 1 
    \end{pmatrix}.
\end{equation}

(ii) The final step is to operate SEB 1 and 2 and measure the charge stability diagram between QDs 3 and 4, resulting in:
\begin{align}
    G_{34} =& - \frac{1-G_{31}G_{13}-G_{32}G_{23}}{r_3^{34}\big|_{(1,2)}} - G_{31}G_{14} - G_{32}G_{24} \\
    G_{43} =& -r_4^{34}\big|_{(1,2)}\left( 1- G_{41}G_{14}-G_{42}G_{24}\right) - G_{41}G_{13} - G_{42}G_{23}.
\end{align}

Therefore, we can fully remove the effect of the PID on the measured slopes and construct the required transformation matrix $\bm{G}$. 

\bibliographystyle{apsrev4-1}
\bibliography{bibliography}

\end{document}